\documentclass[10pt,aps,prd,amsmath,amssymb,twoside,showpacs,nofootinbib,preprintnumbers]{revtex4}
\usepackage[T1]{fontenc}

\newcommand{\defeq}{:=}
\newcommand{\im}{\mathrm{i}}  
\newcommand{\Z}{\mathbb{Z}}
\newcommand{\R}{\mathbb{R}}
\newcommand{\C}{\mathbb{C}}
\newcommand{\xd}{\mathrm{d}}
\newcommand{\xD}{\mathcal{D}} 
\newcommand{\cS}{\mathcal{S}}
\newcommand{\cH}{\mathcal{H}}
\newcommand{\tens}{\otimes}

\newcommand{\be}{\begin{equation}}
\newcommand{\ee}{\end{equation}}

\begin{document}

\title{States and amplitudes for finite regions\\ in a two-dimensional
  Euclidean quantum field theory}
\author{Daniele Colosi}\email{colosi@matmor.unam.mx}
\author{Robert Oeckl}\email{robert@matmor.unam.mx}
\affiliation{Instituto de Matem\'aticas, Universidad Nacional Aut\'onoma
  de M\'exico, Campus Morelia,
C.P.~58190, Morelia, Michoac\'an, Mexico}
\date{23 April 2009 (v2)}
\preprint{UNAM-IM-MOR-2008-2}

\begin{abstract}
We quantize the Helmholtz equation (plus perturbative interactions)
in two dimensions to illustrate a manifestly local description of
quantum field theory. Using the general boundary formulation we 
describe the quantum dynamics both in a traditional time evolution
setting as well as in a setting referring to finite disk (or annulus)
shaped regions of spacetime. We demonstrate that both descriptions are
equivalent when they should be.

\end{abstract}

\maketitle

\section{Introduction}

The general boundary formulation of quantum theory (GBF)
\cite{Oe:boundary,Oe:GBQFT,Oe:probgbf} offers a new way to study
the quantum theory of fields. A main feature of this approach is the
possibility to associate Hilbert spaces of states with arbitrary
hypersurfaces of spacetime. All the information about the physical
processes taking place within a spacetime region
is encoded in the amplitude associated with such a region and
states on its boundary hypersurface.
A key aspect of this approach is the absence of the
requirement of a special type of spacetime hypersurfaces for the
construction of the quantum theory. The GBF should be implementable for
spacetime regions of \emph{arbitrary} form. This peculiar
characteristic of the GBF contrasts dramatically with the conventional
formulation of quantum field theory where a special class of
hypersurfaces is singled out, namely flat spacelike hypersurfaces
defined by a constant value of Minkowskian time. Indeed state spaces
are defined on such equal-time surfaces and transition amplitudes are
defined between two such surfaces. This structure of conventional
quantum field theory is the reflection of the unique role played by
time in quantum theory as the parameter labeling the evolution.

The first extension of 
the standard formulation of quantum field theory 
relevant in the present context
has been introduced by Tomonaga and Schwinger in the late 40s
\cite{Tom:relwave,Sch:qed1}. In their works they
described
the evolution of fields quantized on arbitrary spacelike hypersurfaces
through the so called Tomonaga-Schwinger equation, which generalizes
the functional Schr\"odinger equation. 
Other generalized quantization
prescriptions have also been discussed in the literature, among which we
recall the light-front dynamics \cite{Kom:commchar}, where the surface of
quantization is the plane $t+z= const.$, the covariant formulation of
the light-front dynamics \cite{CDKM:covLFD}, in which the wave
functions are defined on the plane characterized by the equation
$v_{\mu} x^{\mu} =0$ where $v$ is an arbitrary light-like four vector,
and the quantization on the Lorentz invariant spacetime
hyperboloid $x_{\mu} x^{\mu}=const.>0$ \cite{FuHaJa:newft}, also
known as point-form dynamics\footnote{These different forms to
  describe dynamics have been advocated by Dirac in \cite{Dir:forms}
  at the level of classical relativistic dynamics.}. The different
characteristics of these approaches depend on the properties of the
quantization surface, whose exact form is therefore of paramount
importance.

The GBF is compatible with all the above mentioned descriptions of
quantum field theory. Moreover, it should be possible to view them as
special cases.
In the GBF one should be able to describe the dynamics of
quantized fields in spacetime regions whose geometry is different and
even incompatible with the mathematical structures on which the other
mentioned approaches are based. To be more precise, in the different
formalisms
of QFT mentioned above transition
amplitudes are defined for evolution processes that involve an
infinite spacetime region bounded by two disconnected spacelike (or
lightlike)
hypersurfaces. Such a geometry is dictated on the one hand by certain
quantization prescriptions (given by canonical commutation rules to be
imposed on the quantization surface). On the other hand it is imposed
by the standard picture of
dynamics understood as the evolution from an initial state (defined on
an initial spacelike hypersurface) to a final one (defined on a final
spacelike hypersurface) and the associated probability interpretation,
limited to \emph{transition} amplitudes.
But if we are interested in a dynamical
process taking place in a spacetime region naturally bounded by, say, one
connected hypersurface containing timelike parts, the use of the
mentioned forms of QFT may become awkward or even impossible. (As an
extreme case think of a stationary black hole spacetime.)
In contrast,
the GBF can handle such a case at least without conceptual
difficulty. The \emph{technical} problems however, may be considerable in
general.

The conjecture that standard QFT admits an extension in the sense of
the GBF has been addressed in a series of papers:
In \cite{Oe:timelike} it was shown (in the context of free scalar QFT)
that states on certain \emph{timelike} hypersurfaces and amplitudes between
them can be consistently defined and interpreted. A first example of a
region with a timelike and \emph{connected} boundary was given in
\cite{Oe:KGtl}. The region in question is a timelike hypercylinder,
i.e., a ball in space extended over all of time.
Here, due to the connectedness of the boundary,
amplitudes cannot be thought of as transition
amplitudes between different boundary components.
Nevertheless, a consistent probability interpretation was demonstrated.
Perturbatively interacting QFT was treated in
\cite{CoOe:spsmatrix,CoOe:smatrixgbf} showing that an interacting
asymptotic amplitude can be defined from the large radius limit of the
hypercylinder. Moreover, it was demonstrated that this amplitude is
equivalent to the usual S-matrix when both can be defined.

Although the connectedness of the boundary is certainly a novelty, the
hypercylinder shares with the other geometries considered in the
literature so far the property of being non-compact. In the present article
we make a more radical departure from standard QFT and consider
a \emph{finite} region with connected boundary.\footnote{Note that finite
regions in a GBF context have been considered already in the case of
Yang-Mills theory in two dimensions \cite{Oe:2dqym}. However, this theory
is almost topological and thus differs substantially from realistic QFTs.}
In contrast to the other mentioned articles, we operate here in a
setting of \emph{Euclidean} spacetime, as this leads to considerable
technical simplifications.
We study the
quantization of a field theory obeying a Helmholtz equation of motion
and its perturbations in two spacetime dimensions (the generalization
to higher dimensions
presents no conceptual difficulty). Two different geometries are
studied. First, we consider the case of parallel ``equal-time''
hyperplanes with the enclosed region representing
``time-evolution''. This corresponds to a traditional setup,
describing the quantum theory in terms of states evolving in time.
Then, we consider hypersurfaces given by concentric circles in
spacetime and the regions enclosed by them, i.e., discs and annuli of
different radii. The annuli amplitudes can be thought of as describing
the ``evolution'' of states between different radii. In contrast, the
disc amplitudes have no conventional ``transition''
interpretation. The conceptual meaning of the amplitudes is similar to
that in the hypercylinder
setting first described in \cite{Oe:KGtl}.

The plan of this work largely follows that of
\cite{CoOe:smatrixgbf}. We consider in Section
\ref{sec:classicalconfig} the classical field theory in the two
settings mentioned above, and we express the classical solutions of
the equation of motion in terms of the boundary configurations in the
different settings. In Section \ref{sec:quantumtheory} the quantum
theory is presented in the Schr\"odinger representation and the field
propagators for the different regions considered are specified. Vacuum
and coherent states are introduced in Section \ref{sec:vacuumstate}
and \ref{sec:coherent-state} respectively.
Amplitudes for the free theory in both geometries are
evaluated in Section \ref{sec:ampl-free} and their relation studied in
Section \ref{sec:relfreeampl}, where
we construct an isometry between the respective
state spaces which makes the amplitudes equal in the interaction
picture. In Section
\ref{sec:source} we introduce sources to describe perturbative
interactions and construct asymptotic amplitudes of the interacting
theory for both geometries (one of them being essentially the
conventional S-matrix). The case of general interactions is treated in
Section \ref{sec:genint} by means of functional techniques. Finally we
present our conclusion in Section \ref{sec:conclusions}.

Note that, although we work in Euclidean spacetime, our setting is
conceptually distinct from considering the Wick-rotation of a
Lorentzian theory. Rather, we view the theory here as a \emph{Riemannian} real-time QFT
in its own right. In particular, although there are certain formal
similarities, what we are doing is essentially different from the
technique known as \emph{radial quantization} \cite{FuHaJa:newft}, where
one maps a
Lorentzian cylinder to a Euclidean plane. In radial quantization
Hilbert spaces can also be thought of as associated to circles in a
plane, but the latter
are interpreted as images of usual equal-time
hypersurfaces in a Lorentzian cylinder, mapped to the plane.
A consequence of this conceptual difference is, for example, that the
Hilbert space associated in the present setting to a circle
corresponds to a tensor product of \emph{two} equal-time Hilbert
spaces rather than to just one (as would be the case in radial
quantization).

\section{Classical theory}
\label{sec:classicalconfig}

We consider a real massive scalar field in 2 dimensional Euclidean
spacetime which obeys a Helmholtz equation of motion. We will be
interested in studying the field in two different kinds of spacetime
regions. On the one hand we consider an infinite region bounded by two
parallel straight lines. This
represents the traditional point of view on quantum field theory of
evolution between equal time hypersurfaces. On the other hand we
consider a region with the shape of a disk (and also a region with the
shape of an annulus). This is more adapted to
the idea that we want to describe physics locally.

Choosing Cartesian coordinates $\tau,x$ we consider the action
\be
S_{M,0}(\phi)=\frac{1}{2}\int_M \xd \tau \, \xd x\, \left((\partial_{\tau}
\phi)(\partial_{\tau} \phi)+ (\partial_x \phi)(\partial_x \phi) -
m^2\phi^2\right),
\label{eq:freeact}
\ee
for a region $M$ in spacetime. The associated equation of motion is
the Helmholtz equation,
\be
\left( \partial_{\tau}^2 + \partial_x^2 + m^2\right) \phi = 0.
\label{eq:eom}
\ee

\subsection{Slice regions}
\label{sec:standard}

Consider the spacetime region $M=[\tau_1,\tau_2] \times \mathbb{R}$,
where we may think of $[\tau_1,\tau_2]$ as a ``time
interval''. Bounded solutions of (\ref{eq:eom}) in the region $M$ can
be expanded in
Fourier modes in ``space'' and either in Fourier modes or in
exponentials in ``time'',
\be
\phi(\tau,x) = \int_{- \infty}^{+ \infty} \frac{\xd \nu}{2 \pi} \,
\left( a(\nu) e^{\im \omega_\nu \tau} + b(\nu)
e^{-\im \omega_\nu \tau}\right) e^{\im \nu x} .
\ee
Here the ``time'' variable ${\tau}$ belongs to the interval
$[\tau_1,\tau_2]$ and we define
\be
 \omega_\nu \defeq \begin{cases}
 \sqrt{m^2-\nu^2} & \text{if}\quad |\nu|\le m \\
  -\im \sqrt{\nu^2-m^2} & \text{if}\quad |\nu|> m .
 \end{cases}
\ee
For $|\nu|>m$ the field is a
combination of solutions oscillating in space and exponentially
increasing or decreasing in time. For $|\nu| \leq m$, the field
is oscillating both in space and time. The conditions for the field to
be real are $\overline{a(\nu)}=b(-\nu)$ and $\overline{b(\nu)}=
a(-\nu)$ if $|\nu|\le m$. Otherwise they are
$\overline{a(\nu)}=a(-\nu)$ and $\overline{b(\nu)}=b(-\nu)$.

It will be useful in the following to work with solutions of the
equation of motion with specific boundary conditions. In particular a
field $\phi$, obeying equation (\ref{eq:eom}), with boundary
configurations $\varphi_1$ at ${\tau}=\tau_1$ and $\varphi_2$ at
${\tau}=\tau_2$ can be formally written as\footnote{Here and in the following we use the symbol $\phi$ for field configurations in spacetime regions, while we use the symbol $\varphi$ for field configurations on hypersurfaces.}
\be
\phi(\tau,x)= \frac{ \sin \omega(\tau_2 - \tau)}{ \sin \omega(\tau_2
  - \tau_1)} \, \varphi_1(x) +  \frac{ \sin \omega(\tau - \tau_1)}{
  \sin \omega(\tau_2 - \tau_1)} \, \varphi_2(x).
\label{eq:clsol-boundary}
\ee
The symbol $\omega$ denotes the operator whose eigenvalues on a Fourier
expansion in the $x$-direction are the values $\omega_\nu$.

\subsection{Disk and annulus regions}
\label{sec:polar}

We now consider spacetime regions with the shape of a disc and with the
shape of an annulus. For this purpose it is convenient to introduce
polar coordinates $r,\vartheta$ where $\tau=r \sin\vartheta$ and
$x=r \cos\vartheta$. The equation of motion (\ref{eq:eom}) in polar
coordinates is,
\be
\left( \partial_r^2 + \frac{1}{r} \partial_r + \frac{1}{r^2}
\partial^2_{\vartheta} + m^2 \right) \phi=0.
\label{eq:eompolar}
\ee
We expand solutions in terms of Fourier modes around the circle,
\be
\phi(r,\vartheta) = \sum_{n=- \infty}^{+\infty} f_n(r) e^{\im n
  \vartheta}.
\label{eq:fourier}
\ee
Local solutions of (\ref{eq:eompolar}) can then be written in terms of
Bessel functions,
\be
f_n(r) = a_n J_n(m r) + b_n Y_n(m r).
\ee
$J_n$ and $Y_n$ are the
Bessel functions of the first and second kind respectively. The
conditions for the field to be real read $\overline{a_n}=(-1)^n
a_{-n}$ and $\overline{b_n}=(-1)^n b_{-n}$.
Notice
that these Bessel functions have different behavior at the origin
$(r=0)$, 
\be
J_0(0) =1, \qquad J_n(0)=0 \ \  (n=\pm 1, \pm2,...), \qquad \lim_{r
  \rightarrow 0} Y_n(r) = -\infty \  \ \forall n.
\ee

Consider now a disk region of radius $R$ around the origin. Due to the
divergence of the Bessel functions of the second kind at the origin,
only Bessel functions of the first kind are admissible in
solutions. Given boundary data $\varphi(\vartheta)$ at radius $R$ we
can thus formally reconstruct a solution in the whole disk,
\be
\phi(r, \vartheta) = \frac{J_n(m r)}{J_n(m R)} \varphi(\vartheta) .
\label{eq:classol}
\ee
The fraction $\frac{J_n(m r)}{J_n(m R)}$ is to be understood as
an operator, defined through its eigenvalues on a Fourier
decomposition on the circle.

We will be interested also in the annulus region, i.e., the
region bounded by two circles around the origin.
In the annulus region the classical
solution may contain both kinds of Bessel functions because both $J_n$ and
$Y_n$ are regular there. Hence the classical solution can be
decomposed formally in the form
\be
\phi(r, \vartheta) = J_n(m r) \, \varphi_{J}(\vartheta) +
Y_n(m r) \, \varphi_{Y}(\vartheta),
\label{eq:div+conv}
\ee
where $\varphi_J$ and $\varphi_Y$ are real functions on the circle.
$J_n(m r)$ and $Y_n(m r)$ are understood as operators, defined
through their eigenvalues on a Fourier decomposition on the circle.
Denoting
with $\varphi$ and $\hat{\varphi}$ the configurations on the
boundaries of the annulus specified by the circles of radii $R$ and
$\hat{R}$, with $\hat{R} > R$, respectively, we express the classical
solution (\ref{eq:div+conv}) in terms of the boundary configurations
as we did in (\ref{eq:classol}),
\be
\begin{pmatrix}
	\varphi \\
	\hat{\varphi}
\end{pmatrix}
 =
\begin{pmatrix}
	J_n(m R) & Y_n(m R) \\
	J_n(m \hat{R}) & Y_n(m \hat{R})
\end{pmatrix}
\begin{pmatrix}
	\varphi_{J} \\
	{\varphi}_{Y}
\end{pmatrix} .
\ee
Inverting we arrive at
\be
\begin{pmatrix}
	\varphi_{J} \\
	{\varphi}_{Y}
\end{pmatrix}
= 
\frac{1}{\delta_n(mR,m \hat{R})}
\begin{pmatrix}
	Y_n(m \hat{R}) &  -Y_n(m R) \\
	-J_n(m \hat{R}) & J_n(m {R})
\end{pmatrix}
\begin{pmatrix}
	\varphi \\
	\hat{\varphi}
\end{pmatrix},
\ee
where 
\be
\delta_n(z,\hat{z}) \defeq J_n(z) \, Y_n(\hat{z}) - Y_n(z) \, J_n(\hat{z}).
\label{eq:delta}
\ee
Hence, the classical solution in the annulus region takes the form
\be
\phi(r, \vartheta) = \frac{J_n(m r) \, Y_n(m \hat{R}) - Y_n(m r) \,
  J_n(m \hat{R})}{\delta_n(mR,m \hat{R})} \, \varphi(\vartheta) +
\frac{J_n(m R) \, Y_n(m r) - Y_n(m R) \, J_n(m r)}{\delta_n(mR,m
  \hat{R})} \, \hat{\varphi}(\vartheta).
\ee

\section{Quantum theory}
\label{sec:quantumtheory}

The passage to the quantum theory is implemented by the Feynman path
integral prescription, which is the quantization procedure most suited
for the GBF. Moreover, the quantum dynamics of the field is described
in the Schr\"odinger representation, where the quantum states are wave
functionals on the space of field configurations. Thus, we associate
state spaces $\cH_\Sigma$ of wave functions with certain hypersurfaces
$\Sigma$ in spacetime. Amplitudes $\rho_{M}:\cH_{\partial M}\to\C$ are
associated to certain
spacetime regions $M$. (Here $\partial M$ denotes the boundary of $M$.)
State spaces and amplitudes satisfy a number of consistency
conditions, see \cite{Oe:GBQFT} or \cite{Oe:KGtl}.

The amplitude
associated with a region $M$ and a state $\psi$ is given by
\be
\rho_M(\psi) = \int \xd \varphi \, \psi(\varphi) Z_M(\varphi),
\label{eq:amplitude}
\ee
where the integral is extended over all the configurations $\varphi$
on the boundary of the region $M$. $Z_M(\varphi)$ is the field
propagator, formally defined as
\be
Z_M(\varphi) = \int_{\phi|_{\partial M}=\varphi} \xD\phi\, e^{\im S_{M}(\phi)},
\label{eq:proppint0}
\ee
where $S_M(\phi)$ is the action of the field in $M$. In the
case of the free theory determined by the free action
(\ref{eq:freeact}) we can
evaluate the associated propagator $Z_{M,0}$ by shifting the
integration variable by a classical solution matching the boundary
configuration $\varphi$ in $\partial M$. Explicitly,
\be
Z_{M,0}(\varphi) = \int_{\phi|_{\partial M}=\varphi} \xD\phi\, e^{\im S_{M,0}(\phi)}
=  \int_{\phi|_{\partial M}=0} \xD\phi\, e^{\im
  S_{M,0}(\phi_{cl}+\phi)} = N_{M,0} \,  e^{\im
  S_{M,0}(\phi_\text{cl})} ,
\label{eq:propfree}
\ee
where the normalization factor is formally given by
\be
N_{M,0}=\int_{\phi|_{\partial M}=0} \xD\phi\, e^{\im S_{M,0}(\phi)}.
\label{eq:n00}
\ee

In order to identify states and amplitudes explicitly, it is
convenient to proceed in the following order: 1. Work out amplitudes
and verify their gluing properties. 2. Identify the vacuum state for
each relevant hypersurface. 3. Identify particle states or other
relevant states. In the following we will carry this out both for the
slice regions of Section~\ref{sec:standard} and for the
disc or annulus type regions of Section~\ref{sec:polar}.

\subsection{Propagator in the slice region}
\label{sec:propinf}

We evaluate the propagator associated with the
slice spacetime region
$M=[\tau_1,\tau_2] \times \mathbb{R}$. (In the formulas below
  we indicate such a region with the subscript
  $[\tau_1,\tau_2]$.) The boundary $\partial M$ is the disjoint
union of the lines $\tau=\tau_1$ and $\tau = \tau_2$. The boundary
state $\psi(\varphi)$ results to be the product of the wave function
$\psi_1(\varphi_1)$, describing the state of the system at "time"
$\tau_1$, with the wave function $\psi_2(\varphi_2)$, describing the
state of the system at "time" $\tau_2$. By (\ref{eq:propfree}) we may
use the classical solution
(\ref{eq:clsol-boundary}) of the equation of motion that interpolates
between $\varphi_1$ at $\tau_1$ and $\varphi_2$ at $\tau_2$ in order
to express the field propagator of the free theory. The result is,
\be
 Z_{[\tau_1,\tau_2],0}(\varphi_1,\varphi_2)=N_{[\tau_1,\tau_2],0}
 \exp\left(\frac{\im}{2}\int\xd x\,
 \begin{pmatrix}\varphi_1 & \varphi_2\end{pmatrix}
   W_{[\tau_1,\tau_2]}  \begin{pmatrix}\varphi_1\\ \varphi_2\end{pmatrix}\right),
 \label{eq:freeprop}
\ee
where the normalization factor is formally given by
\be
N_{[\tau_1,\tau_2],0}=\int_{\substack{\phi|_{\tau_1}=0\\ \phi|_{\tau_2}=0}}
\xD\phi\, e^{\im S_{[\tau_1,\tau_2],0}(\phi)},
\label{eq:n0}
\ee
and $W_{[\tau_1,\tau_2]}$ is the operator valued $2\times 2$ matrix
\be
W_{[\tau_1,\tau_2]}=\frac{\omega}{\sin \omega(\tau_2-\tau_1)}
\begin{pmatrix} \cos \omega(\tau_2-\tau_1) & -1\\
 -1 & \cos \omega(\tau_2-\tau_1)\end{pmatrix}.
\label{Matrix}
\ee
The relevant gluing property in this context is the composition of two
``temporally'' consecutive slice regions. We do not write down
this calculation explicitly here.

\subsection{Propagator in the disk and annulus regions}
\label{sec:propfin}
 
As already mentioned, we are interested in two types of regions here: The disk
region of radius $R$, indicated with label $R$, and the annulus
region, bounded by
two circles of radii $R$ and $\hat{R}$ (we assume $R<\hat{R}$),
indicated with label $[R,\hat{R}]$. For the disk region the field
propagator is evaluated using (\ref{eq:propfree}) with the classical
solution (\ref{eq:classol}),
\be
Z_{R,0}(\varphi) = N_{R,0} \, \exp \left( \frac{\im}{2} \int \xd
\vartheta \, m R \, \varphi(\vartheta) \, \frac{J_n'(m R)}{J_n(m R)}
\,  \varphi(\vartheta) \right).
\label{propR}
\ee
For the annulus region we use the classical solution
(\ref{eq:div+conv}) and obtain for the propagator the expression
\be
Z_{[R,\hat{R}],0}(\varphi,\hat{\varphi}) = N_{[R,\hat{R}],0}  \, \exp
\left( \frac{\im}{2} \int \xd \vartheta \, 
\begin{pmatrix}\varphi & \hat{\varphi} \end{pmatrix} W_{[R,\hat{R}]}  
\begin{pmatrix} \varphi \\ \hat{\varphi} \end{pmatrix}\right),
\label{propRR}
\ee
with
\be
W_{[R,\hat{R}]} = \frac{m}{\delta_n(mR,m \hat{R})}
\begin{pmatrix} R \sigma_n(m \hat{R}, mR) & - \frac{2}{\pi m}\\
 - \frac{2}{\pi m} & \hat{R} \sigma_n (mR,m\hat{R}) \end{pmatrix}.
\label{W}
\ee
The function $\delta_n$ has been defined in (\ref{eq:delta}). The function $\sigma_n$ is to be understood as the operator defined as
\begin{equation}
\sigma_n(\hat{z},z)  \defeq J_n(\hat{z}) \, Y_n'(z) - J_n'(z) \, Y_n(\hat{z}).
\end{equation}

It can be shown that the propagators (\ref{propR}) and (\ref{propRR})
satisfy the following composition rules,
\be
Z_{\hat{R},0}(\hat{\varphi}) = \int \xD \varphi \, Z_{R,0}(\varphi)
\, Z_{[R,\hat{R}],0}(\varphi,\hat{\varphi}), 
\ee
and, for $R_1<R_2<R_3$,
\be
Z_{[R_1,R_3],0}(\varphi_1,\varphi_3) = \int \xD \varphi_2 \,
Z_{[R_1,R_2],0}(\varphi_1,\varphi_2)\,
Z_{[R_2,R_3],0}(\varphi_2,\varphi_3).
\ee
These relations prove the consistency of the definitions (\ref{propR})
and (\ref{propRR}).

\section{Vacuum state}
\label{sec:vacuumstate}

The vacuum state has to satisfy the vacuum axioms \cite{Oe:GBQFT}, see
also \cite{Oe:KGtl}. This implies in particular that it is invariant
under ``evolution''. In the case of the vacuum on an ``equal time''
line this means ``time evolution'' along a slice region. In
the case of the vacuum on the circle this means (generalized)
invariance under radial evolution along an annulus region.

As in \cite{Oe:timelike,Oe:KGtl} we make for the vacuum wave function
on a hypersurface $\Sigma$ the
Gaussian ansatz
\be
 \psi_{\Sigma,0}(\varphi)=C  \exp\left(-\frac{1}{2}\int_\Sigma \xd
 x\,\varphi(x)(A \varphi)(x)\right) ,
\ee
where $C$ is a normalization factor and $A$ an unknown operator.

\subsection{Constant $\tau$ line}

In the first case we want to determine the vacuum on hypersurfaces
that are lines of constant $\tau$.
To determine the operator $A$ we
consider the free evolution from $\tau_1$ to $\tau_2$ encoded in the
propagator (\ref{eq:freeprop}). The invariance of the vacuum state can
be written in the following form
\be
\psi_{0}(\varphi_2) = \int \xD \varphi_1 \,
\psi_{0}(\varphi_1)  \,
Z_{[\tau_1,\tau_2],0}(\varphi_1,\varphi_2).
\ee
This equation implies for the operator $A$ to satisfy $A^2=\omega^2$.
We select the solution $A=\omega$ so that $A$ be bounded from
below. Hence, the vacuum state can be written in momentum space as
\be
\psi_{0}(\varphi)=C  \exp\left( -\frac{1}{2} \int_{-\infty}^{+\infty} \frac{\xd
  \nu}{2 \pi} \, \varphi(\nu)  \omega_{\nu}  \varphi(-\nu)\right).
\ee
The normalization factor $C$ is then fixed (up to a phase),
\be
|C|^{-2}= \int \xD \varphi \exp\left( -\frac{1}{2} \int_{- m}^m
\frac{\xd \nu}{2 \pi} \, \varphi(\nu) 2 \omega_{\nu} \,
\varphi(-\nu)\right).
\ee

\subsection{Circle}

For the vacuum state defined on the circle of radius $R$, the operator
$A \equiv A_R$ denotes a family of operators indexed by the radius
$R$. Demanding generalized invariance of the vacuum state under evolution,
namely
\be
\psi_{R,0}({\varphi}) = \int \xD \hat{\varphi} \,
\psi_{\hat{R},0}(\hat{\varphi})  \,
Z_{[R,\hat{R}],0}(\varphi,\hat{\varphi})
\ee
leads to an equation for the operator $A_R$,
\be
\left( \im m \sigma_n (m\hat{R},mR) + \delta_n(mR,m \hat{R}) A_R
\right) \left(-\im m \sigma_n (mR,m\hat{R}) + \delta_n(mR,m
\hat{R}) A_{\hat{R}} \right) =\frac{4}{\pi^2 R\hat{R}}.
\label{eq:B}
\ee
The solutions of the above equation are of the form
\be
A_R = \im m \frac{\mathcal{C}_n'}{\mathcal{C}_n},
\label{eq:vacuum-op}
\ee
where $\mathcal{C}_n$ can be one of the following cylindrical
functions: Bessel functions of the first and second kind, $J_n$ and
$Y_n$, or Hankel functions of the first and second kind, defined
respectively as $H_n= J_n +\im Y_n$ and $\overline{H}_n = J_n - \im
Y_n$.
If we require that the argument of the exponential in the vacuum state
be bounded from below, we must select the Hankel function of the
second kind as the cylindrical function in
(\ref{eq:vacuum-op}).
The vacuum state can then be written as
\be
\psi_{R,0}(\varphi)= C_R \exp \left( - \frac{\im}{2} \int \xd
\vartheta \, \varphi(\vartheta) \left( m R
\frac{\overline{H}_n'(mR)}{\overline{H}_n(mR)}  \varphi
\right)(\vartheta) \right).
\label{eq:vacuum2}
\ee
The normalization factor is given (up to a phase) by the equation
\be
|C_R|^{-2} = \int \xD \varphi \, \exp \left( - \frac{2}{\pi} \int \xd
\vartheta \, \varphi(\vartheta) \left( \frac{1}{|H_n(mR)|^2}  \varphi
\right)(\vartheta)  \right)
\ee

\subsection{Comparison between the vacuum states}

In order to compare the vacuum states in the two settings, we express
the operators that define these two vacua, namely $\omega$ and $A_R$
obtained in the preceding sections, in an appropriate asymptotic
region. In particular we consider a small region near the positive
$\tau$ axis at large radius $R$. In polar coordinates this means we take
$\vartheta \approx \pi /2$ at $r=R$. Hence we have $\partial_x \approx
\frac{1}{R} \partial_{\vartheta}$. Then, the operator $\omega$
in this asymptotic region now reads
\be
\omega \approx  \sqrt{\frac{\partial_{\vartheta}^2}{R^2} + m^2}.
\ee
On the Fourier expansion (\ref{eq:fourier}) this yields the eigenvalues
$\sqrt{-\frac{n^2}{R^2}+m^2}$. Fixing $n$ we take the limit
$R\rightarrow\infty$ to obtain $\omega \rightarrow m$ in this sense. With
the asymptotic expansion of the Bessel functions $J_n$ and $Y_n$ (see
\cite{AbSt:handbook} for example) it is easy to evaluate the
expression of $A_R$ for large
radius: $A_R \rightarrow m$. Hence, in this asymptotic sense,
$\omega \approx A_R$. With another choice in (\ref{eq:vacuum-op})
this would not be
true. The freedom we found in
selecting the vacuum on a constant $\tau$ line is indeed linked in
this way to the freedom in selecting the vacuum on the circle.

\section{Coherent states}
\label{sec:coherent-state}

In \cite{CoOe:spsmatrix,CoOe:smatrixgbf} coherent states have been an
essential tool for the computation of asymptotic amplitudes. This is
equally the case in the present paper. To define coherent states we
use an approach parallel to the one in \cite{CoOe:smatrixgbf}.

\subsection{Constant $\tau$ line}

We define a coherent state at constant $\tau$ in the Fock representation as
\be
|\psi_{\eta}\rangle =  C_{\eta} \exp \left( \int_{- m}^m
\frac{\xd \nu}{2 \pi} \frac{1}{2 \omega_{\nu}} \, \eta(\nu)
a^{\dagger}(\nu) \right) |0\rangle,
\label{cs}
\ee
where $a^{\dagger}(\nu)$ is the creation operator associated with the
mode $\nu$ of the field, $|0\rangle$ represents the vacuum state and
$\eta(\nu)$ is a complex function on the interval $[-m,m]$.
Note that the restriction of $\nu$ to this interval is analogous to
the restriction in \cite{CoOe:smatrixgbf} of the coherent states on
the hypercylinder to depend only on physical configurations. The
meaning of ``physical configurations'' in the present context is that
those are the modes that behave well (i.e., are bounded) for large
$\tau$.
The normalization factor is
\be
C_{\eta} = \exp \left( - \frac{1}{2} \int_{- m}^m \frac{\xd \nu}{2 \pi} \frac{1}{2 \omega_{\nu}} |\eta(\nu)|^2  \right).
\ee
In the Schr\"odinger representation the coherent state reads
\be
\psi_{\eta}(\varphi) =  K_{\eta} \exp \left( \int_{- m}^m
\frac{\xd \nu}{2 \pi} \, \eta(\nu) \varphi(\nu) \right)
\psi_0(\varphi),
\ee
where 
\be
K_{\eta} = \exp \left( - \frac{1}{2} \int_{- m}^m \frac{\xd
  \nu}{2 \pi} \frac{1}{2 \omega_{\nu}} \left( \eta(\nu) \eta(-\nu) +
|\eta(\nu)|^2 \right) \right).
\ee
The characteristic property of coherent states is to remain coherent
under the action of the free propagator: Coherent states evolve to
coherent states,
\be
\psi_{\eta_2}(\varphi_2) = \int \xD \varphi_1 \,
\psi_{\eta_1}(\varphi_1)  \,
Z_{[\tau_1,\tau_2],0}(\varphi_1,\varphi_2).
\ee
This equation yields the following relation for the complex functions
$\eta_1$, defined on the surface $\tau=\tau_1$ and $\eta_2$, defined
on the surface $\tau=\tau_2$,
\be
\eta_2(\nu) = \eta_1(\nu) e^{- \im \omega_{\nu} (\tau_2 - \tau_1)}.
\ee
In the interaction picture a coherent state can thus be defined as
\be
\psi_{\tau,\eta}(\varphi) =  K_{\tau,\eta} \exp \left( \int_{- m}^m \frac{\xd \nu}{2 \pi} \, \eta(\nu) e^{- \im \omega_{\nu}
  \tau} \varphi(\nu) \right) \psi_0(\varphi),
\label{csint}
\ee
where the normalization factor $K_{\tau,\eta}$ is now time dependent,
\be
K_{\tau,\eta} = \exp \left( - \frac{1}{2} \int_{- m}^m
\frac{\xd \nu}{2 \pi} \frac{1}{2 \omega_{\nu}} \left( e^{- 2 \im
  \omega_{\nu} \tau} \eta(\nu) \eta(-\nu) + |\eta(\nu)|^2 \right)
\right).
\label{eq:normint}
\ee

It will be useful to expand the coherent state (\ref{cs}) in terms of multiparticle states,
\be
| \psi_{\eta} \rangle = C_{\eta} \sum_{n=0}^{\infty} \frac{1}{n!} \int_{- m}^m \frac{\xd \nu_1}{2 \pi 2 \omega_{\nu_1}} \cdots \frac{\xd \nu_n}{2 \pi 2 \omega_{\nu_n}}\,  \eta(\nu_1) \cdots \eta(\nu_n) \, | \psi_{\nu_1, \dots, \nu_n} \rangle.
\label{eq:n-part1}
\ee
The inner product between a coherent state defined by the complex function $\eta$ and a state with $n$ particles of quantum numbers $\nu_1, \dots, \nu_n$ is,
\be
\langle  \psi_{\nu_1, \dots, \nu_n} | \psi_{\eta} \rangle = C_{\eta}  \eta(\nu_1) \cdots \eta(\nu_n).
\label{scprcoh1}
\ee
The inner product of two $n$-particle states is
\be
\langle  \psi_{\nu_1, \dots, \nu_n} |  \psi_{\mu_1, \dots, \mu_n} \rangle = \frac{1}{n!} \sum_{\sigma \in S_n} \prod_{i=1}^n 2 \pi 2 \omega_{\nu_i} \delta(\nu_i-\mu_{\sigma(i)}).
\ee
The sum runs over all permutations $\sigma$ of $n$ elements.

\subsection{Circle}

We define the coherent states on the circle of radius $R$ in terms of
complex coefficients $\eta_n$ as 
\be
\psi_{R,\eta} (\varphi)= K_{R,\eta} \exp \left( \sum_n \eta_n
\varphi_n \right) \psi_{R,0}(\varphi),
\ee
where the normalization factor is 
\be
K_{R,\eta} = \exp \left( - \sum_n \frac{|H_n(m R)|^2}{ 16} (\eta_n
\eta_{-n} + |\eta_n|^2) \right).
\ee
Coherent states remain coherent under free propagation, namely
\be
\psi_{R,\eta} (\varphi) = \int \xD \hat{\varphi} \,
\psi_{\hat{R},\hat{\eta}} (\hat{\varphi}) \,
Z_{[R,\hat{R}],0}(\varphi,\hat{\varphi}).
\ee
This equation is satisfied provided that the complex functions $\eta$
at radius $R$ and $\hat{\eta}$ at radius $\hat{R}$ are related by
\be
\eta_n = \hat{\eta}_n \, \frac{\overline{H}_n(m
  \hat{R})}{\overline{H}_n(m R)}.
\ee
Then the interaction picture can be defined with the coefficients
$\xi_n = \overline{H}_n(m R) \eta_n$, and a coherent state takes the
form
\be
\psi_{R,\xi} (\varphi)= K_{R,\xi} \exp \left( \sum_n
\frac{\xi_n}{\overline{H}_n(m R)} \, \varphi_n \right)
\psi_{R,0}(\varphi).
\ee
The normalization factor is then
\be
K_{R,\xi} = \exp \left( - \frac{1}{16} \sum_n \left[ \frac{H_n(m
    R)}{\overline{H}_{-n}(m R)} \xi_n \xi_{-n} + |\xi_n|^2 \right]
\right).
\label{eq:normfac}
\ee
The coherent states satisfy the following completeness relation,
\be
D^{-1} \int \xd \xi \, \xd \overline{\xi} \, |\psi_{\xi} \rangle \langle \psi_{\xi}| = I,
\label{eq:complrel}
\ee
where $I$ is the identity operator and the constant $D$ has the form
\be
D = \int \xd \xi \xd \overline{\xi} \exp \left( - \sum_n \frac{|\xi_n|^2}{8}\right).
\label{Dtilde}
\ee
The $k$-particle expansion of the coherent state determined by the complex function $\xi$ reads
\be
 | \psi_{\xi}\rangle = \exp \left(- \sum_n \frac{|\xi_n|^2}{16} \right)\sum_{k=0}^{\infty} \frac{1}{k!} \sum_{n_1} \dots \sum_{n_k} \xi_{n_1} \cdots \xi_{n_k} |\psi_{n_1, \dots, n_k}\rangle,
\label{eq:n-part2}
\ee
where $|\psi_{n_1, \dots, n_k}\rangle$ denotes the state with $k$ particles of quantum numbers $n_1, \cdots, n_k$. The inner product between coherent state $\psi_{\xi}$ and an $k$-particle state then results to be
\be
\langle \psi_{n_1, \dots, n_k} | \psi_{\xi}\rangle = \exp \left(- \sum_n \frac{|\xi_n|^2}{16} \right) \frac{\xi_{n_1}}{8} \cdots \frac{\xi_{n_k}}{8}.
\label{scprcoh2}
\ee
The inner product between two $k$-particle states is 
\be
\langle \psi_{n_1, \dots, n_k} | \psi_{n'_1, \dots, n'_k}\rangle = \frac{8^{-k}}{k!} \sum_{\sigma \in S_k} \prod_{i=1}^k \delta_{n_{\sigma(i)},n'_i},
\ee
where the sum is over all the permutations $\sigma$ of $k$ elements.

\section{Amplitudes in the free theory}
\label{sec:ampl-free}

We compute in the following the amplitudes of coherent states in the theory
determined by the free action (\ref{eq:freeact}). We use the
interaction picture so that amplitudes will be independent of
``elapsed time'' $\tau$ or radius $r$. In particular, we may think of
the amplitudes obtained as asymptotic amplitudes.

\subsection{Slice region}

The transition amplitude from the coherent state defined by the
complex function $\eta_1$ on the hypersurface $\tau=\tau_1$ to the
coherent state defined by $\eta_2$ on the hypersurface $\tau=\tau_2$
in the interaction picture is given by
\be
\rho_{[\tau_1,\tau_2],0}( \psi_{\tau_1,\eta_1} \otimes
\overline{\psi_{\tau_2,\eta_2}})
=\int \xD\varphi_1\,\xD\varphi_2 \, \psi_{\tau_1,\eta_1}(\varphi_1)
\overline{\psi_{\tau_2,\eta_2} (\varphi_2)}\,
Z_{[\tau_1,\tau_2],0}(\varphi_1,\varphi_2).
\label{eq:tampl-free}
\ee
The above expression reduces to the expression of the inner product
between two coherent states defined by the complex functions $\eta_1$
and $\eta_2$,
\be
\langle \psi_{\eta_2}|\psi_{\eta_1}\rangle = \exp \bigg( \int_{-m}^m
\frac{\xd \nu}{(2 \pi) 2 \omega_{\nu}} \, \left(  \eta_1(\nu)\,
\overline{\eta_2(\nu)}  - \frac{1}{2} |\eta_1(\nu)|^2  - \frac{1}{2}
|\eta_2(\nu)|^2\right) \bigg),
\label{eq:sproduct-cs}
\ee
and is therefore independent of the initial and final times $\tau_1$
and $\tau_2$. We can view this as the S-matrix of the free theory,
sending $\tau_1 \rightarrow - \infty$ and
$\tau_2 \rightarrow + \infty$,
\be
\cS_0 ( \psi_{\eta_1} \otimes \overline{\psi_{\eta_2}}) 
= \lim_{\substack{\tau_1\to-\infty\\ \tau_2\to+\infty}}
\rho_{[\tau_1,\tau_2],0}( \psi_{\tau_1,\eta_1} \otimes
\overline{\psi_{\tau_2,\eta_2}})
=\langle \psi_{\eta_2}|\psi_{\eta_1}\rangle .
\label{eq:smfree}
\ee

\subsection{Disk region}

The amplitude associated with a coherent state in the interaction
picture in the disk region is
\begin{align}
\rho_{R,0}(\psi_{R,\xi}) &=  \int \xD \varphi \,
\psi_{R,\xi}(\varphi) Z_{R,0}(\varphi), \nonumber\\
&= N_{R,0}K_{R,\xi} C_R \int \xD \varphi \, \exp \left(  \sum_n
\left[ \frac{\xi_n}{\overline{H}_n(m R)} \varphi_n - \varphi_n
  \frac{2}{\overline{H}_n(mR) J_n(m R)} \varphi_{-n} \right] \right).
\end{align}
We shift the integration variable by the quantity
\be
\Delta_{n} = \frac{J_{-n}(m R)}{4} \xi_{-n},
\ee
and we obtain the amplitude
\be
\rho_{R,0}(\psi_{R,\xi})= \exp \left(  \sum_n \frac{1}{16} ((-1)^n\xi_n
\xi_{-n} - |\xi_n|^2 ) \right).
\label{eq:freeampl}
\ee
Notice that this amplitude is independent of the radius $R$, as it
should be by construction. The limit $R \rightarrow \infty$ is then
trivial, and the asymptotic amplitude, $\cS_0$, is therefore
\be
\cS_0(\psi_{\xi}) = \lim_{R \rightarrow \infty}
\rho_{R,0}(\psi_{R,\xi})= \exp \left(  \sum_n \frac{1}{16} ((-1)^n\xi_n
\xi_{-n} - |\xi_n|^2 ) \right).
\label{eq:asamplfree}
\ee

\section{Relation between states and amplitudes in the two settings}
\label{sec:relfreeampl}

So far we have kept the description of the two settings (slice
regions with line boundaries versus disk/annulus regions with circle
boundaries) completely separate. However, the objects (states and
amplitudes) that we have constructed in the two settings should be
compatible with each other. There is indeed a way to ensure this
compatibility. Consider a slice region $[\tau_1,\tau_2]\times\R$
where we cut out a disk of radius $R$ around the origin. (Assume
$\tau_1<-R$ and $\tau_2> R$.) Suppose we work out the amplitude for
this ``punched'' region. Compatibility then means that: (a) the
composition of the amplitude of a punched region with the amplitude of
a disk fitting into the hole yields the amplitude of the resulting
slice region and (b) the composition of the amplitude of a
punched region with the amplitude of 
an annulus region fitting into the hole yields the amplitude of the
resulting punched region.

A direct calculation of the amplitude of a punched region in the way
we have done the calculations for the other types of regions would
be very complicated and we will not attempt it here. We will,
however, be able to infer this amplitude indirectly. It is a map
$\rho_{[\tau_1,\tau_2,R]}:\cH_1\tens\cH_2^*\tens\cH_R^*\to\C$. Here,
$\cH_1$ denotes the Hilbert space associated with the line
$\tau=\tau_1$, with its orientation being induced by it being in the
boundary of the punched region. $\cH_2$ denotes 
the Hilbert space associated with the line $\tau=\tau_2$, with the
same (translated) orientation, i.e., its orientation is opposite to
the one induced by it being in the boundary of the punched
region. $\cH_R$ is the state space of the circle, oriented as the
boundary of a disk. Thinking about the classical boundary value
problem we should expect the induced map
$\tilde{\rho}_{[\tau_1,\tau_2,R]}:\cH_1\tens\cH_2^*\to \cH_R$ to be an
isomorphism. (In the terminology of \cite{Oe:GBQFT} the state
spaces $\cH_1$ and $\cH_2$ are of size $1/2$ while the state space
$\cH_R$ is of size $1$).

The concept of such an isomorphism is familiar from
\cite{CoOe:spsmatrix,CoOe:smatrixgbf}, where a similar situation
occurs. There, an isomorphism between the tensor product of an
initial and final state space (similar to $\cH_1\tens\cH_2^*$
here) and the state space on a hypercylinder (similar to $\cH_R$
here) is constructed in Klein-Gordon theory. In that case there is no
region that
interpolates between the two types of hypersurface and thus no
interpolating amplitude.\footnote{If one introduces
  ``negative'' regions then it is possible to consider such amplitudes.
But this is another story on which we will not expand here.} Instead,
the isomorphism emerges from a comparison of amplitudes for the theory
coupled to a source. We will follow the same route in the present
work.

Nevertheless, we introduce the isomorphism already here since it is
valid whether or not we include sources. Only the justification for
choosing the isomorphism precisely in this way may seem weak in the
present context. We shall see later on, in the context of the theory
with source that this choice is forced upon
us. An alternative way to justify the exact form of the isomorphism
could be to perform a semiclassical analysis (which suggests itself
because we are using coherent states). However, we will not do this
here.

Let $\eta_1$ and $\eta_2$ be complex functions on the interval $[-m,m]$
determining coherent states $\psi_{\tau_1,\eta_1}\in\cH_1$ and
$\overline{\psi_{\tau_2,\eta_2}}\in\cH_2^*$. We define the following complex
solution of the Helmholtz equation in spacetime,
\be
\hat{\eta}(\tau,x)= \int_{-m}^m \frac{\xd \nu}{2\pi} \frac{1}{2 \omega_{\nu}}
\left(\eta_1(\nu) e^{-\im \omega_{\nu} \tau - \im \nu x}
+\overline{\eta_2(\nu)} e^{\im \omega_{\nu} \tau + \im \nu x} \right).
\label{eq:classcoh}
\ee
This establishes a one-to-one correspondence between bounded
complex classical solutions in spacetime and coherent states in
$\cH_1\tens\cH_2^*$.
Let $\{\xi_n\}_{n\in\Z}$ be complex coefficients defining a coherent
state $\psi_{R,\xi}\in\cH_R$. We define the following complex solution
of the Helmholtz equation in spacetime,
\be
\hat{\xi}(r,\vartheta)= \frac{1}{4} \sum_n \xi_n \, J_n(m r)  e^{- \im
  n \vartheta}.
\label{eq:xihat}
\ee
This establishes a one-to-one correspondence between bounded
complex classical solutions in spacetime and coherent states in
$\cH_R$.

Now that we have two correspondences between coherent states and
classical solutions we use these to identify the coherent states in
$\cH_1\tens\cH_2^*$ with those in $\cH_R$. That is, for a complex
classical solution $\hat{\zeta}$ in spacetime we compute on the one
hand its components $\zeta_1$ and $\zeta_2$ in terms of
(\ref{eq:classcoh}) and on the other hand the coefficients $\zeta_n$
in terms of (\ref{eq:xihat}). The 
isomorphism $\cH_1\tens\cH_2^*\to\cH_R$ is then determined by
$\psi_{\tau_1,\zeta_1}\tens\overline{\psi_{\tau_2,\zeta_2}} \mapsto
\psi_{R,\zeta}$.

\subsection{Equivalence of amplitudes}

We proceed to show that the isomorphism between state spaces does indeed
lead to an equivalence between the amplitude for a slice
region and the amplitude for a disk region. This amounts to
demonstrating the gluing property of the amplitude of
the punched region with the amplitude of the disk region. Again, we
do not need to fix  $\tau_1$, $\tau_2$ or
$R$ since we use the interaction picture. Rather, for convenience, we
can think of amplitudes as asymptotic
amplitudes, i.e., the expressions (\ref{eq:smfree}) and
(\ref{eq:asamplfree}).
Explicitly, we are going to show that,
\be
\cS_0 ( \psi_{\zeta_1} \otimes \overline{\psi_{\zeta_2}})
= \cS_0 ( \psi_{\zeta}).
\ee
We start by expressing the classical solution $\hat{\zeta}$ in terms of
Bessel functions of the first kind. We introduce in
(\ref{eq:classcoh}) the variable $\alpha = \arcsin \frac{\nu}{m}$,
\be
\hat{\zeta}(\tau,x) = \int_{-\pi/2}^{\pi/2} \frac{\xd \alpha}{4 \pi}
\left( \zeta_1(m \sin \alpha) e^{-\im m \tau \cos \alpha - \im m x \sin
  \alpha} + \overline{\zeta_2}(m \sin \alpha) e^{\im m \tau \cos \alpha
  + \im m x \sin \alpha} \right).
\ee
In the polar coordinates this expression takes the form
\be
\hat{\zeta}(r,\vartheta) = \int_{-\pi/2}^{\pi/2} \frac{\xd \alpha}{4
  \pi} \left( \zeta_1(m \sin \alpha) e^{-\im m r \sin (\alpha+
  \vartheta)} + \overline{\zeta_2}(m \sin \alpha) e^{\im m r \sin
  (\alpha+ \vartheta)} \right).
\ee
We rewrite the exponentials in terms of the Bessel function of the
first kind as
\be
e^{\pm \im m r \sin (\alpha + \vartheta)} = \sum_n e^{\im n(\alpha +
  \vartheta)} J_n( \pm m r). 
\ee
Hence,
\be
\hat{\zeta}(r,\vartheta) = \frac{1}{4} \sum_n J_n(m r) e^{\im n
  \vartheta} \int_{-\pi/2}^{\pi/2} \frac{\xd \alpha}{\pi} e^{\im n
  \alpha} \left( \zeta_1(m \sin \alpha)  + (-1)^n \, \overline{\zeta_2}(m
\sin \alpha)  \right).
\ee
Under the identification of the complex classical solutions of the
Helmholtz equation (\ref{eq:classcoh}) and (\ref{eq:xihat}) we obtain,
\be
\zeta_{-n} = \int_{-\pi/2}^{\pi/2} \frac{\xd \alpha}{\pi} e^{\im n
  \alpha} \left( \zeta_1(m \sin \alpha)  + (-1)^n \, \overline{\zeta_2}(m
\sin \alpha)  \right).
\label{xi-eta}
\ee
Substituting this in the free amplitude (\ref{eq:asamplfree}),
\begin{align}
\cS_{0}(\psi_{\zeta}) 
= & \exp \left(  \sum_n  \left[ (-1)^n 
  \int \frac{ \xd \alpha \, \xd \alpha'}{16 \pi^2} \, e^{\im n (\alpha
    - \alpha')} \left( \zeta_1(m \sin \alpha)  + (-1)^n \,
  \overline{\zeta_2}(m \sin \alpha)  \right) \left( \zeta_1(m \sin
  \alpha')  + (-1)^{-n} \, \overline{\zeta_2}(m \sin \alpha')  \right)
  \right. \right. \nonumber\\
  & - \left. \left.      \int \frac{ \xd \alpha \, \xd \alpha'}{\pi^2}
  \, e^{\im n (\alpha - \alpha')}   \left( \zeta_1(m \sin \alpha)  +
  (-1)^n \, \overline{\zeta_2}(m \sin \alpha)  \right)\left(
  \overline{\zeta_1}(m \sin \alpha')  + (-1)^n \, \zeta_2(m \sin \alpha')
  \right)  \right] \right), \nonumber
\end{align}
Notice that 
\be
\int_{- \pi /2}^{\pi / 2} \xd \alpha \int_{- \pi /2}^{\pi / 2} \xd
\alpha' \, \delta(\alpha- \alpha' + \pi) =0 .
\ee
We arrive at
\be
\cS_{0}(\psi_{\zeta}) = \exp \left(   \int_{-\pi/2}^{\pi/2}
 \frac{ \xd \alpha}{4
    \pi} \left(  \zeta_1(m \sin \alpha)\overline{\zeta_2}(m \sin \alpha)
  - \frac{1}{2}\zeta_1(m \sin \alpha) \overline{\zeta_1}(m \sin \alpha)
  - \frac{1}{2}\zeta_2(m \sin \alpha) \overline{\zeta_2}(m \sin \alpha)
  \right) \right).
\ee
Substituting $\alpha = \arcsin \frac{\nu}{m}$, we obtain
\be
\cS_{0}(\psi_{\zeta}) 
=  \exp \left(  \int_{-m}^m \frac{\xd \nu}{2 \pi}  \,
  \frac{1}{2 \omega_\nu} \left[ \zeta_1( \nu) \, \overline{\zeta_2( \nu)}  -
  \frac{1}{2} |\zeta_1(\nu)|^2 - \frac{1}{2} |\zeta_2( \nu)|^2\right]
\right) = \langle \psi_{\zeta_2} | \psi_{\zeta_1} \rangle
= \cS_0 ( \psi_{\zeta_1} \otimes \overline{\psi_{\zeta_2}}).
\ee
This concludes the proof of the equivalence of the free amplitudes
(\ref{eq:smfree}) and (\ref{eq:asamplfree}).

\subsection{Isomorphism in terms of multiparticle states}

We turn to work out the form of isomorphism
$\cH_1\tens\cH_2^*\to\cH_R$ in terms of multiparticle
states. Note that we are using the interaction picture and hence
may omit the labels $\tau_1$, $\tau_2$ and $R$.
We denote a state in $\cH_1 \otimes \cH_2^*$ with $q$ incoming
particles with quantum numbers $\nu_1, \dots, \nu_q$ and $k-q$
outgoing particles with quantum numbers $\nu_{q+1}, \dots, \nu_k$ as
\be
\psi_{\nu_1, \dots, \nu_q | \nu_{q+1}, \dots, \nu_k} = |\psi_{\nu_{1},
  \dots, \nu_q} \rangle \otimes \langle  \psi_{\nu_{q+1}, \dots,
  \nu_k} |,
\ee
where $|\psi_{\nu_{1}, \dots, \nu_q} \rangle$ is a $q$-particle state
in $\cH_1$ introduced in (\ref{eq:n-part1}). The scalar product of
this $k$-particle state, $\psi_{\nu_1, \dots, \nu_q | \nu_{q+1},
  \dots, \nu_k}$, with a coherent state defined by the complex
function $\hat{\zeta}$ results to be
\be
\langle \psi_{\hat{\zeta}} | \psi_{\nu_1, \dots, \nu_q | \nu_{q+1},
  \dots, \nu_k} \rangle = \langle \psi_{\zeta_1} | \psi_{\nu_1, \dots,
  \nu_q} \rangle \langle \psi_{\nu_{q+1}, \dots, \nu_k} |
\psi_{\zeta_2} \rangle =  \overline{C_{\zeta_1}} C_{\zeta_2}
\zeta_2(\nu_1) \cdots \zeta_2(\nu_q) \overline{\zeta_1}(\nu_{q+1})
\cdots \overline{\zeta_1}(\nu_k),
\label{scprcoh}
\ee
where the relation (\ref{scprcoh1}) has been used. With the use of the
correspondence (\ref{xi-eta}), we can express this inner product in
terms of the modes $\zeta_n$. From (\ref{xi-eta}) we derive the
following relations,
\be
\zeta_1(\nu) = \sum_{n} \frac{\zeta_{n}}{2} e^{\im
  n \kappa_\nu}, \qquad
\overline{\zeta_2}(\nu) = \sum_{n} (-1)^{n}\frac{\zeta_{n}}{2}
e^{\im n \kappa_\nu},
\ee
where $\kappa_\nu\defeq \arctan \frac{\nu}{\omega_\nu}$.
Inserting these expressions in (\ref{scprcoh}) yields,
\be
\langle \psi_{\hat{\zeta}} | \psi_{\nu_1, \dots, \nu_q | \nu_{q+1},
  \dots, \nu_k} \rangle = \exp \left(- \sum_n \frac{|\zeta_n|^2}{16}
\right) \sum_{n_1, \dots, n_k} (-1)^{n_1+ \cdots +
  n_q} \frac{\overline{\zeta_{n_1}}}{2}
\cdots \frac{\overline{\zeta_{n_k}}}{2} \, e^{-\im
  \left(n_1 \kappa_{\nu_1}+\cdots
 +n_k \kappa_{\nu_k}\right)}.
\label{ip1}
\ee
With the identification of the complex classical solutions
(\ref{eq:classcoh}) and (\ref{eq:xihat}) and the completeness relation
of the coherent states (\ref{eq:complrel}), 
we are now able to express the inner product of a $k$-particle state
in $\cH_1 \otimes \cH_2^*$ with a $k$-particle state in $\cH_{R}$,
\be
\langle \psi_{n_1, \dots, n_k} |\psi_{\nu_1, \dots, \nu_q | \nu_{q+1},
  \dots, \nu_k} \rangle = {D}^{-1} \int \xd \hat{\zeta} \xd
\overline{\hat{\zeta}} \langle \psi_{n_1, \dots, n_k} |
\psi_{\hat{\zeta}}\rangle \langle \psi_{\hat{\zeta}} | \psi_{\nu_1,
  \dots, \nu_q | \nu_{q+1}, \dots, \nu_k} \rangle,
\label{ip3}
\ee
Inserting (\ref{Dtilde}), (\ref{scprcoh2}), (\ref{ip1}) in (\ref{ip3})
and performing the integration in $\xd \hat{\zeta} \xd
\overline{\hat{\zeta}}$, we obtain
\be
\langle \psi_{n_1, \dots, n_k} |\psi_{\nu_1, \dots, \nu_q | \nu_{q+1},
  \dots, \nu_k} \rangle =  (-1)^{n_1+ \cdots + n_q} \, 2^{-k} \,
e^{-\im \left(n_1 \kappa_{\nu_1} + \cdots + n_k \kappa_{\nu_n} \right)} \frac{1}{k!} \sum_{\sigma \in S_k} \prod_{i=1}^k \delta_{n_1,n_{\sigma(i)}'},
\ee
where the sum runs over all permutations $\sigma$ of $k$ elements. Hence a $k$-particle state in $\cH_1 \otimes \cH_2^*$ can be written
as a linear combination of $k$-particle states in $\cH_{R}$ as,
\be
\psi_{\nu_1, \dots, \nu_q | \nu_{q+1}, \dots, \nu_k} = \sum_{n_1,
  \cdots, n_k} (-1)^{n_1+ \cdots + n_q} \, 4^{k} \, e^{-\im \left(n_1
  \kappa_{\nu_1} + \cdots + n_k \kappa_{\nu_k}\right)} \psi_{n_1, \dots, n_k}.
\ee
Reciprocally, a $k$-particle state in $\cH_{R}$ is a linear
combination of $k$-particle states in $\cH_1 \otimes \cH_2^*$,
\be
\psi_{n_1, \dots, n_k} =  \int \frac{\xd \nu_1}{2 \pi 2
  \omega_{\nu_1}} \cdots \frac{\xd \nu_k}{2 \pi 2 \omega_{\nu_k}}\,
(-1)^{{n_{1}}+ \cdots + {n_{q}}} \, 2^{-k} \, e^{-\im \left({n_{1}}
  \kappa_{\nu_1} + \cdots + {n_{k}} \kappa_{\nu_k} \right)}
\psi_{\nu_1, \dots, \nu_q | \nu_{q+1}, \dots, \nu_k}.
\ee

\section{Asymptotic amplitudes in theory with source}
\label{sec:source}

We study in this section the interaction of the field with a source
field $\mu$ described by the action
\be
S_{M,\mu}(\phi)=S_{M,0}(\phi) + \int_M \xd \tau\,\xd x\,
 \phi(\tau,x) \mu(\tau,x),
\label{eq:actionsrc}
\ee
where $S_{M,0}$ is the free action (\ref{eq:freeact}). The resulting
field propagator can be expressed in terms of the one of the free
theory. Indeed in the expression of the path integral defining the
propagator, we shift the integration variable by a classical solution
of the free theory matching the boundary configurations $\varphi$ on
the boundary $\partial M$, and we obtain the result
\be
Z_{M,\mu}(\varphi)= \frac{N_{M,\mu}}{N_{M,0}} Z_{M,0}(\varphi)
\exp\left( \im \int_M \xd \tau \, \xd x\, \phi_\text{cl}(\tau,x)
\mu(\tau,x)\right),
 \label{eq:propsrc2}
\ee
where the normalization factor $N_{M,\mu}$ is formally equal to
\be
N_{M,\mu}=\int_{\phi|_{\partial M}=0} \xD\phi\, e^{\im S_{M,\mu}(\phi)}.
\label{eq:nmu}
\ee
In order to relate this normalization factor to that of the free
theory $N_{M,0}$ (\ref{eq:n00}), we shift the integration variable in
(\ref{eq:nmu}) by the function $\alpha$, solution of the inhomogeneous
Helmholtz equation
\be
\left( \partial_{\tau}^2 + \partial_x^2 + m^2\right) \alpha(\tau,x)=\mu(\tau,x),
\label{eq:iKG1}
\ee
with the boundary condition $\alpha \big|_{\partial M}=0$. We can now
rewrite the normalization factor (\ref{eq:nmu}) as
\be
N_{M,\mu} = N_{M,0} \exp\left(\frac{\im}{2}\int_M \xd \tau\, \xd
x\,\mu(\tau, x)\alpha(\tau, x)\right).
\label{eq:alphafac}
\ee

\subsection{Slice region}
\label{sec:ta-source}

We assume that the source field $\mu$ vanishes outside the slice
region $[\tau_1,\tau_2] \times \mathbb{R}$. We then rewrite
the last exponential in (\ref{eq:propsrc2}) using
(\ref{eq:clsol-boundary}) as,
\be
\exp\left( \im \int\xd \tau \, \xd x\, \phi_\text{cl}(\tau,x)
\mu(\tau,x)\right) =  \exp\left(\int\xd x\,\left(\varphi_1(x)\mu_1(x)
 +\varphi_2(x)\mu_2(x)\right)\right),
\ee
where $\varphi_i$ is the field configuration at $\tau_i$ and
\be
 \mu_1(x) \defeq  \im \int_{\tau_1}^{\tau_2}\xd \tau \, \frac{ \sin
   \omega(\tau_2 - \tau)}{ \sin \omega(\tau_2 - \tau_1)} \mu(\tau,x),
 \qquad
\mu_2(x) \defeq \im \int_{\tau_1}^{\tau_2}\xd \tau \, \frac{ \sin
  \omega(\tau - \tau_1)}{ \sin \omega(\tau_2 - \tau_1)} \mu(\tau,x)
\label{eq:mu}.
\ee
The amplitude $\rho_{[\tau_1,\tau_2],\mu}$ associated with the
transition from the coherent state $\psi_{\tau_1,\eta_1}$ at
$\tau=\tau_1$ to
the coherent state $\psi_{\tau_2,\eta_2}$ at $\tau=\tau_2$ is, in the
interaction picture,
\begin{multline}
\rho_{[\tau_1,\tau_2],\mu}(\psi_{\tau_1,\eta_1} \otimes
\overline{\psi_{\tau_2,\eta_2}}) =
K_{\tau_1,\eta_1}\overline{K_{\tau_2,\eta_2}} \int
\xD\varphi_1\,\xD\varphi_2
\exp \left( \int_{-m}^{m} \frac{\xd \nu}{2 \pi} \, \left(\eta_1(\nu)
 \, e^{- \im \omega_{\nu} \tau_1} \, \varphi_1(\nu)+\overline{\eta_2(-\nu)}
\, e^{\im \omega_{\nu} \tau_2} \, \varphi_2(\nu) \right)\right)\\
\psi_0(\varphi_1)\,\overline{\psi_0(\varphi_2)}\,
Z_{[\tau_1,\tau_2],\mu}(\varphi_1,\varphi_2) .
\label{eq:tampl-src2}
\end{multline}
Introducing two new complex functions $\tilde{\eta}_1$ and
$\tilde{\eta}_2$ defined as
\be
\tilde{\eta}_1(\nu)\defeq \eta_1(\nu) + \int \xd x\, e^{\im
  \omega_{\nu} \tau_1+ \im \nu x} \mu_1(x) \quad \text{and} \quad
\tilde{\eta}_2(\nu)\defeq \eta_2(\nu) + \int \xd x\, e^{ \im
  \omega_{\nu} \tau_2+ \im \nu x} \overline{\mu_2}(x),
\ee
we can rewrite (\ref{eq:tampl-src2}) in the form
\be
\rho_{[\tau_1,\tau_2],\mu}(\psi_{\tau_1,\eta_1} \otimes
\overline{\psi_{\tau_2,\eta_2}})  =
\rho_{[\tau_1,\tau_2],0}(\psi_{\tau_1,\tilde{\eta}_1} \otimes
\overline{\psi_{\tau_2,\tilde{\eta}_2}}) \,
\frac{N_{[\tau_1,\tau_2],\mu}\, K_{\tau_1,\eta_1}\,
  \overline{K_{\tau_2,\eta_2}}}{N_{[\tau_1,\tau_2],0}\,
  K_{\tau_1,\tilde{\eta}_1}\,
  \overline{K_{\tau_2,\tilde{\eta}_2}}}.
\label{eq:tampl-src3}
\ee
Substituting the expressions of the inner product
(\ref{eq:sproduct-cs}) and the normalization factors
(\ref{eq:normint}), we obtain
\begin{multline}
\rho_{[\tau_1,\tau_2],\mu}(\psi_{\tau_1,\eta_1} \otimes
\overline{\psi_{\tau_2,\eta_2}}) =
\rho_{[\tau_1,\tau_2],0}(\psi_{\tau_1,\eta_1} \otimes
\overline{\psi_{\tau_2,\eta_2}})  \,
\frac{N_{[\tau_1,\tau_2],\mu}}{N_{[\tau_1,\tau_2],0}} \exp\left(\im
\int\xd \tau\, \xd x\, \mu(\tau,x) \hat{\eta}(\tau,x)\right) \cdot \\
\exp\left(\int \frac{\xd x}{4 \omega} \left( \mu_1^2(x) + \mu_2^2(x) +
2 \mu_1(x)  e^{-\im \omega(\tau_2-\tau_1)}\mu_2(x) \right) \right),
\label{eq:tampl-src4}
\end{multline}
where the complex function $\hat{\eta}$ is the complex classical
solution of the Helmholtz equation determined by the $\eta_1$ and
$\eta_2$ introduced in (\ref{eq:classcoh}).
Substituting the expressions of the function $\mu_1$ and $\mu_2$ given
in (\ref{eq:mu}), the last exponential in (\ref{eq:tampl-src4}) can be
written in the form
\be
\exp\left(\int \frac{\xd x}{4 \omega} \left( \mu_1^2(x) + \mu_2^2(x) +
2 \mu_1(x) e^{-\im \omega(\tau_2-\tau_1)}\mu_2(x) \right) \right) =
\exp\left( \frac{\im}{2} \int \xd \tau\, \xd x\,\mu(\tau, x)\beta(\tau,
x)\right),
\label{eq:betafac}
\ee
where $\beta$ is the solution of the Helmholtz equation given by
\be
\beta(\tau, x) = \int_{\tau_1}^{\tau_2} \frac{\xd \tau'}{2 \omega}
 \left( \im e^{\im
  \omega (\tau - \tau')} + 2\frac{\sin (\omega(\tau-\tau_1)) \, \sin
  (\omega(\tau_2-\tau'))}{\sin (\omega(\tau_2-\tau_1))} \right)
\mu(\tau',x).
\label{eq:beta}
\ee

We now turn to the quotient of normalization factors
$\frac{N_{[\tau_1,\tau_2],\mu}}{N_{[\tau_1,\tau_2],0}}$ appearing in
(\ref{eq:tampl-src4}). It can be expressed using (\ref{eq:alphafac}) as
\be
 \frac{N_{[\tau_1,\tau_2],\mu}}{N_{[\tau_1,\tau_2],0}}=\exp\left(\frac{\im}{2}\int
 \xd \tau \xd x\,\mu(\tau, x)\alpha(\tau, x)\right),
\label{eq:alphafac1}
\ee
where $\alpha$ is a solution of equation (\ref{eq:iKG1}) with the
boundary conditions $\alpha(\tau_1,x)=0$ and $\alpha(\tau_2,x)=0$. $\alpha$
results to be
\be
\alpha(\tau,x) =   \int_{\tau_1}^{\tau_2} \frac{\xd \tau'}{\omega}
\left( \theta(\tau- \tau') \,  \sin (\omega (\tau- \tau')) -
\frac{\sin (\omega (\tau-\tau_1)) \, \sin (\omega
  (\tau_2-\tau'))}{\sin (\omega (\tau_2-\tau_1))} \right) \,
\mu(\tau',x),
\label{eq:alpha}
\ee
where $\theta(t)$ is the step function
\be
\theta(t)= \begin{cases} 1 & \ \ \text{if} \ \ t>0\\
0 & \ \ \text{if} \ \ t<0.
\end{cases}
\ee
Summing the functions $\alpha(\tau,x)$ given by (\ref{eq:alpha}) and
$\beta(\tau,x)$ given by (\ref{eq:beta}),
\be
\gamma(\tau,x) := \alpha(\tau,x)+ \beta(\tau,x) =
\int_{\tau_1}^{\tau_2} \frac{\xd
  \tau'}{\omega}  \left( \theta(\tau- \tau') \,  \sin (\omega (\tau-
\tau')) + \frac{\im}{2} e^{\im \omega (\tau- \tau')} \right) \,
\mu(\tau',x),
\ee
we obtain a negative frequency solution of the inhomogeneous Helmholtz
equation for $\tau < \tau_1$,
\be
\gamma(\tau,x)\bigg|_{\tau < \tau_1}  = \int_{\tau_1}^{\tau_2}
 \frac{\xd \tau'}{\omega}
\frac{\im}{2} e^{\im \omega (\tau- \tau')}\, \mu(\tau',x),
\label{bc0}
\ee
and a positive frequency solution for $\tau > \tau_2$,
\be
\gamma(\tau,x)\bigg|_{\tau > \tau_2}  = \int_{\tau_1}^{\tau_2}
 \frac{\xd \tau'}{\omega}
\frac{\im}{2} e^{-\im \omega (\tau- \tau')}\, \mu(\tau',x).
\label{bc1}
\ee
We combine the factors (\ref{eq:betafac}) and (\ref{eq:alphafac}) to obtain
\be
\exp \left(\frac{\im}{2} \int \xd \tau\, \xd x\, \mu(\tau,x)
\left[\alpha(\tau,x) + \beta(\tau,x) \right]\right) =
\exp \left(\frac{\im}{2} \int \xd \tau\,\xd x\, \xd \tau'\,\xd x'\,
\mu(\tau,x) G(\tau,x, \tau',x') \mu(\tau',x')\right),
\ee
with
\begin{align}
G(\tau,x, \tau',x') &= \im \int_{-\infty}^{\infty} \frac{\xd \nu}{2
  \pi} \left[ \theta(\tau - \tau') \frac{e^{ \im \omega_{\nu}
      (\tau'-\tau) - \im \nu (x-x')}}{2 \omega_{\nu}} + \theta(\tau' -
  \tau) \frac{e^{- \im \omega_{\nu} (\tau'-\tau)- \im \nu (x-x')}}{2
    \omega_{\nu}}  \right], \nonumber\\
&= \im \int_{0}^{\infty} \frac{\xd \nu}{2 \pi} \frac{e^{ - \im
    \omega_{\nu} |\tau'-\tau|}}{ \omega_{\nu}} \cos (\nu(x-x')).
\end{align}
We recognize on the right-hand side the integral representation of the
Hankel function \cite{GR:tables}. Introducing the 2-dimensional
vectors $\underline{r}$ and $\underline{r}'$, with components
$(\tau,x)$ and $(\tau',x')$ respectively, the function $G(\tau,x,
\tau',x')$ takes the form
\be
G(\underline{r}, \underline{r}') =  \frac{\im}{4} \overline{H_0}(m
|\underline{r} - \underline{r}'|),
\label{eq:green}
\ee
where $H_0$ denotes the Hankel function of order 0.
Hence the transition amplitude now reads
\be
\rho_{[\tau_1,\tau_2],\mu}(\psi_{\tau_1,\eta_1} \otimes
\overline{\psi_{\tau_2,\eta_2}})  =
\rho_{[\tau_1,\tau_2],0}(\psi_{\tau_1,\eta_1} \otimes
\overline{\psi_{\tau_2,\eta_2}})  \,
\exp\left( \im \int\xd^2 x\, \mu(x) \hat{\eta}(x)\right) \exp
\left(\frac{\im}{2} \int \xd^2 x \xd^2 x' \mu(x) G(x, x')
\mu(x')\right).
\label{eq:tampl-src5}
\ee
In order to interpret this transition amplitude as an element of the
S-matrix, denoted by $\cS_{\mu}$, we take the (trivial) limit $\tau_1
\rightarrow -\infty$ and $\tau_2 \rightarrow +\infty$,
\begin{align}
\cS_{\mu}(\psi_{\eta_1} \otimes \overline{\psi_{\eta_2}}) &=
\lim_{\substack{\tau_1\to-\infty\\ \tau_2\to+\infty}}
\rho_{[\tau_1,\tau_2],\mu}(\psi_{\tau_1,\eta_1} \otimes
\overline{\psi_{\tau_2,\eta_2}}), \nonumber\\
&= \cS_{0}(\psi_{\eta_1} \otimes \overline{\psi_{\eta_2}}) \,
\exp\left( \im \int\xd^2 x\, \mu(x) \hat{\eta}(x)\right) \exp
\left(\frac{\im}{2} \int \xd^2 x \xd^2 x' \mu(x) G(x, x')
\mu(x')\right).
\label{eq:smsrc}
\end{align}

\subsection{Disk region}

We now consider a source field $\mu$ vanishing outside the disk
region. As in the preceding section we start by evaluating the last
exponential in (\ref{eq:propsrc2}). With the classical solution
(\ref{eq:classol}), the argument of the exponential reads,
\be
 \int \xd r \, \xd \vartheta \,r \, \mu(r,\vartheta)
 \phi_{cl}(r,\vartheta) = \int \xd r \, \xd \vartheta \, r \,  \mu(r,
 \vartheta) \frac{J_n(m r)}{J_n(m R)} \varphi(\vartheta) =  \sum_n
 \varphi_{-n} \frac{2 \pi}{J_n(m R)}\, j_{n},
\ee
with the quantity $j_n$ given by
\be
j_n \defeq \int \xd r \, r \, J_{n}(m r) \mu_n(r).
\ee
The amplitude of a coherent state in the theory with source is
\be
\rho_{R,\mu}(\psi_{R,\xi}) = \frac{N_{R,\mu}}{N_{R,0}} \int \xD
\varphi \, \psi_{R,\xi}(\varphi) \exp \left( \im \sum_n \varphi_{-n}
\frac{2 \pi }{J_n(m R)}\, j_{n} \right) Z_{R,0}(\varphi).
\ee
The integration can be performed by introducing a new coherent state
defined by the complex function $\tilde{\xi}$ related to $\xi$ via
\be
\tilde{\xi}_n \defeq \xi_n + \im 2 \pi
 \frac{\overline{H}_n(mR)}{J_{-n}(m R)} j_{-n}.
\ee
Then the amplitude has the form
\be
\rho_{R,\mu}(\psi_{R,\xi}) = \frac{N_{R,\mu}}{N_{R,0}}
\frac{K_{R,\xi}}{K_{R,\tilde{\xi}}} \,
\rho_{R,0}(\psi_{R,\tilde{\xi}}).
\ee
The substitution of the expression for the free amplitude
(\ref{eq:freeampl}) of the coherent state defined by $\tilde{\xi}$ and
the expression of the normalization factors, (\ref{eq:normfac}) gives
\be
\rho_{R,\mu}(\psi_{R,\xi}) = \rho_{R,0}(\psi_{R,\xi})
\frac{N_{R,\mu}}{N_{R,0}} \exp \left(  \sum_n \left( \im \frac{\pi}{2}
\xi_n j_n - \frac{\pi^2}{2}j_n \frac{\overline{H}_{-n}(m R)}{J_{n}(m
  R)} j_{-n} \right) \right).
\label{eq:amplsource}
\ee
The first term in the argument of the exponential can be written in
position space as
\be
\exp \left( \im \frac{\pi}{2} \sum_n  \xi_n j_n \right)  = \exp \left(
\im \int \xd^2 x \, \mu(x) \, \hat{\xi}(x)\right),
\ee
where the function $\hat{\xi}$ in polar coordinates is given by
(\ref{eq:xihat}). 
We express the second term in the exponential of (\ref{eq:amplsource})
in the form
\be
\exp \left( -\frac{\pi^2}{2} \sum_n j_n \frac{\overline{H}_n(m
  R)}{J_{-n}(m R)} j_{-n} \right)= \exp \left( \frac{\im}{2}\int \xd^2
x  \, \mu(x) \, \beta(x) \right),
\label{eq:betafactor}
\ee
with $\beta$ given by its Fourier components
\be
\beta_n(r) =  \im \frac{\pi}{2} J_{n}(m r)  \frac{\overline{H}_n(m
  R)}{J_{n}(m R)} j_{n}.
\label{betan}
\ee
We now consider the quotient $\frac{N_{R,\mu}}{N_{R,0}}$. This factor
can be expressed using (\ref{eq:alphafac}) as, 
\be
 \frac{N_{R,\mu}}{N_{R,0}}=\exp\left(\frac{\im}{2}\int \xd^2 x  \,
 \mu(x)\alpha(x)\right),
\label{eq:alphafac-disk}
\ee
where the function $\alpha$ now satisfies the inhomogeneous Helmholtz
equation in polar coordinates (\ref{eq:eompolar}) with the boundary
condition $\alpha(R,\vartheta) = 0$.
It will be convenient to work in momentum space: We consider the
Fourier components of $\alpha$ and $\mu$,
\be
\alpha(r,\vartheta) =  \sum_n \, \alpha_n(r)\, e^{\im n \vartheta},
\quad \mu(r,\vartheta) =  \sum_n \, \mu_n(r)\, e^{ \im n \vartheta}.
\ee
The inhomogeneous Helmholtz equation takes the form
\be
\left( \partial_{r}^2 + \frac{1}{r} \partial_{r} + \left(m^2 -
\frac{n^2}{r^2} \right) \right) \alpha_n(r) = \mu_n(r). 
\ee
The solution is
\be
\alpha_n(r) =\im \frac{\pi}{2} \left( J_n(m r) \left[ h_n(r) - h_n +
  \frac{H_n(mR)}{J_n(mR)} j_n \right] - H_n(m r) \, j_n(r)  \right),
\label{alphan}
\ee
where
\begin{gather}
h_{n}(r) \defeq \int_0^{r} \xd s \, s \, H_n(m s) \mu_{n}(s), \\
h_{n} \defeq \int_0^{\infty} \xd s \, s \, H_n(m s) \mu_{n}(s), \\
j_{n}(r) \defeq \int_0^{r} \xd s \, s \, J_n(m s) \mu_{n}(s). 
\end{gather}
Summing the functions $\alpha_n(r)$ given by (\ref{alphan}) and
$\beta_n(r)$ given by (\ref{betan}),
\be
\gamma_n(r) := \alpha_n(r) + \beta_n(r) = \im \frac{\pi}{2} \left(
J_n(m r) \left[ h_n(r) - h_n + 2 j_n \right] - H_n(m r) \, j_n(r)
\right),
\label{bc2}
\ee
we obtain a solution of the inhomogeneous Helmholtz equation with the
following behavior outside the disk region,
\be
\gamma_n(r)\big|_{r > R} = \im \frac{\pi}{2} \, \overline{H}_n(mr) j_n.
\label{bc3}
\ee
So, combining the factor (\ref{eq:alphafac-disk}) and
(\ref{eq:betafactor}) we arrive at
\be
\exp\left(\frac{\im}{2}\int \xd^2 x  \, \mu(x)[\alpha(x) +
  \beta(x)]\right) = \exp\left(\frac{\im}{2}\int \xd^2 x  \,\xd^2 x'
\, \mu(x)G(x,x')\mu(x')\right),
\ee
where $G$ is the Green function given in polar coordinates by
\be
G(r,\vartheta,r',\vartheta') =  -\frac{\im}{4} \sum_n \, e^{\im n
  (\vartheta-\vartheta')} \left(\theta(r-r')J_n(m r') H_n(m r) +
\theta(r'-r)J_n(m r) H_n(m r')  - 2  J_{n}(m r') J_{n}(m r)\right).
\label{eq:G}
\ee
Using the addition theorems of the Bessel functions (11.3.4) and
(11.3.5) of \cite{Wat:Bessel}, we can perform the sum over $n$ in
(\ref{eq:G}); we obtain
\be
G(r,\vartheta,r',\vartheta') 
= \frac{\im}{4} \, \overline{H}_0(m \sqrt{r^2 + r'^2 - 2 r r' \cos
  (\vartheta-\vartheta')}) .
\ee
If we note by $\underline{r}$ the vector with polar coordinates $(r,
\vartheta)$, the Green function can be written as in equation
(\ref{eq:green}).
The amplitude of a coherent state in presence of a source is then
\be
\rho_{R,\mu}(\psi_{R,\xi}) = \rho_{R,0}(\psi_{R,\xi}) \exp \left( \im
\int \xd^2 x  \, \mu(x) \, \hat{\xi}(x)
\right)\exp\left(\frac{\im}{2}\int \xd^2 x  \,\xd^2 x' \,
\mu(x)G(x,x')\mu(x')\right).
\ee
This expression is independent of the radius $R$ (since we are working
in the interaction picture). The asymptotic
amplitude, $\cS_{\mu}$, is then immediately obtained,
\be
\cS_{\mu}(\psi_{\xi}) = \lim_{R \rightarrow \infty}
\rho_{R,\mu}(\psi_{R,\xi}) = \cS_{0}(\psi_{\xi}) \exp \left( \im \int
\xd^2 x  \, \mu(x) \, \hat{\xi}(x) \right)\exp\left(\frac{\im}{2}\int
\xd^2 x  \,\xd^2 x' \, \mu(x)G(x,x')\mu(x')\right).
\label{eq:asyamplsrc}
\ee

\subsection{Comparison of amplitudes with source}

Recall from Section~\ref{sec:relfreeampl} that there is an isomorphism
between the state space $\cH_1\tens\cH_2^*$
on the boundary of a slice region $[\tau_1,\tau_2]\times\R$
and the state space $\cH_R$ on the boundary of a disk region $S^2_R$
in the free theory. Moreover, under this isomorphism the amplitude of
the slice region and the disk region become equal. We can
extend such a comparison now to amplitudes for the theory coupled to a
source. To this end we consider the asymptotic
amplitudes for $\tau_1\rightarrow -\infty$, $\tau_2\rightarrow
+\infty$ and $R\rightarrow \infty$. (Alternatively, we could compare
amplitudes for finite time intervals and radii as long as the source is
confined completely inside the regions under consideration.) Indeed,
comparing (\ref{eq:smsrc}) with (\ref{eq:asyamplsrc}) we observe
that the two expressions become equal under the isomorphism of
Section~\ref{sec:relfreeampl}: The first factor in both expressions is
the free amplitude which was already shown to be equal. The second
factor in both expressions involves complex classical solutions of the
Helmholtz equation. It were precisely these complex classical
solutions that we used to define the isomorphism. The third factor is
obviously equal in both expressions.

In Section~\ref{sec:relfreeampl} the choice of the isomorphism
$\cH_1\tens\cH_2^*\to\cH_R$ seemed somewhat ad hoc. At this point it
it becomes clear that there is no other choice. In order for the
amplitudes (\ref{eq:smsrc}) and (\ref{eq:asyamplsrc}) to be equal we
need precisely that the complex classical solutions $\hat{\eta}$ and
$\hat{\xi}$ that appear in the expressions for these amplitudes be equal.

\section{General interactions}
\label{sec:genint}

To describe general perturbative interactions we use the usual
technique of functional derivatives with respect to the source
field. 
Thus consider the action
\be
S_{M,V}(\phi)=S_{M,0}(\phi)+\int_M \xd^2 x\, V(x,\phi(x)),
\label{eq:actionpot}
\ee
where $S_{M,0}$ is the free action (\ref{eq:freeact}) and $V$ a potential.
We notice the usual functional identity,
\be
\exp\left(\im S_{M,V}(\phi)\right)= \exp\left(\im\int_M \xd x^2\,
V\left(x,-\im \frac{\partial}{\partial
\mu(x)}\right)\right) \exp\left(\im
S_{M,\mu}(\phi)\right)\bigg|_{\mu=0},
\label{eq:actionfunc}
\ee
where $S_{M,\mu}$ is the action in the presence of a source interaction,
defined in (\ref{eq:actionsrc}). At first we assume the source to
vanish outside the region $M$. We then notice that we can perform all
the calculations of Section~\ref{sec:source} with the action
(\ref{eq:actionpot}) by always pulling out to the left the factor in
(\ref{eq:actionfunc}) with the functional derivative. This finally
leads to the interacting S-matrix in functional form. For the
asymptotic slice regions this is,
\be
  \cS_V(\psi_{1} \otimes \overline{\psi_2} )
 = \exp\left(\im\int\xd^2 x\,
  V\left(x,-\im\frac{\partial}{\partial \mu(x)}\right)\right)\\
 \cS_\mu (\psi_{1} \otimes \overline{\psi_2} )
 \bigg|_{\mu=0} ,
\label{eq:smint}
\ee
while for the asymptotic disk region this is,
\be
\cS_{V}(\psi) =  \exp\left(\im\int\xd^2 x\,
V\left(x,-\im\frac{\partial}{\partial \mu(x)}\right)\right)
\cS_{\mu}(\psi)\bigg|_{\mu=0} .
\label{eq:smint2}
\ee

\section{Conclusions}
\label{sec:conclusions}

We have presented the Riemannian quantum theory of a field obeying Helmholtz's
equation of motion in two-dimensional Euclidean spacetime within the
general boundary formulation in two
different settings. The first setting is conceptually identical to what
is usually done in standard QFT: The state space is defined on a
hypersurface of constant time and the evolution of states
is considered from one such hypersurface to another. On
the other hand, the second setting we studied is incompatible with
standard QFT methods of describing the dynamics of quantized
fields. The novelty here consists of dealing with a compact spacetime
region, the disk region, bounded by one closed line, the circle. We
have shown that this second way to describe the quantum theory is
completely compatible with the first one: The physical predictions of
the two treatments are indeed the same due to the equivalence of the
asymptotic amplitudes both for the free and the general interacting
theory. This equivalence relies on the existence of an isomorphism
between the state spaces defined in the two settings.

Transition amplitudes in the ``traditional'' setting of slice regions
are unitary as should be expected.\footnote{We emphasize again that we
are working in a Riemannian real-time setting and not in a Wick rotated
setting.} Less conventionally, radial ``translation''
amplitudes between circles in the disk/annulus region
setting are also unitary. This means that quantum mechanically
probabilities are conserved under ``radial evolution''.\footnote{See
  \cite{Oe:GBQFT} for the appropriately generalized notion of
probability conservation applicable here.}
Indeed, this should be expected from the
classical field theory. Solutions of the Helmholtz equation in a
finite region have a unique continuation beyond that region. Thus,
``radial evolution'' is well defined classically. The field dynamics
in a smaller and a larger disk are entirely equivalent.
We have thus shown that this is also true quantum
mechanically. Note that this is analogous to what happens in
Klein-Gordon theory for hypercylinders in Minkowski space
\cite{Oe:KGtl,CoOe:spsmatrix,CoOe:smatrixgbf}.

The relevance of the result presented here is that the formulation of
the theory in the disk region implements a fully local description of
the quantum dynamics of the field. Moreover, one can view this as a
kind of finite spacetime holography: The dynamics in a finite spacetime
region is completely described through states on the region's
boundary. Any physical interaction between the region and its
spacetime surroundings factors through the boundary state space.

An important next step will be the realization of general boundary
amplitudes and state spaces for finite regions in a Lorentzian quantum
field theory in Minkoswki space. While we expect such a description to
be feasible, it involves technical challenges related to the fact that
the boundary of such a region would have spacelike as well as timelike
parts. Furthermore, the solutions of classical field equations in
finite spacetime regions no longer determine unique continuations
outside such regions. We expect this classical fact to be reflected
in the quantum theory in that amplitudes corresponding to annulus like
regions no longer permit a representation as unitary operators between
the inner and outer boundary state spaces.

\begin{acknowledgments}

This work was supported in part by CONACyT grants 47857 and 49093.

\end{acknowledgments}

\bibliographystyle{amsordx}
\bibliography{stdrefs}

\end{document}